\documentclass[conference]{IEEEtran}

\usepackage{basicreq}
\usepackage{hyperref}

\def\L{\Lambda}

\title{A  Lattice Coding Scheme for Secret Key Generation from Gaussian Markov Tree Sources}
\author{Shashank Vatedka and Navin Kashyap\\Department of Electrical Communication Engineering\\Indian Institute of Science, Bengaluru, India\\
email: \texttt{\{shashank,nkashyap\}@ece.iisc.ernet.in}}

\begin{document}
\maketitle
\begin{abstract}
In this article, we study the problem of secret key generation in the multiterminal source model, where the terminals have access to correlated Gaussian sources. We assume that the sources form a Markov chain on a tree. We give a nested lattice-based key generation scheme whose computational complexity is polynomial in the number, $N$, of independent and identically distributed samples observed by each source. We also compute the achievable secret key rate and give a class of examples where our scheme is optimal in the fine quantization limit. However, we also give examples that show that our scheme is not always optimal in the limit of fine quantization.
\end{abstract}

\section{Introduction}\label{sec:introduction}
We study secret key (SK) generation in the multiterminal source model, where $m$ terminals possess correlated Gaussian sources.
Each terminal observes $N$ independent and identically distributed (iid) samples of its source. The terminals have access to a noiseless public channel of infinite capacity, and their objective is to agree upon a 
secret key by communicating across the public channel. The key must be such that any eavesdropper having access to the public communication must not be able to guess the key. In other words, the key must be independent (or almost independent) of the messages communicated across the channel. A measure of performance is the secret key rate that can be achieved, which is the number of bits of secret key generated per (source) sample. On the other hand, the probability that any terminal is unable to reconstruct the key correctly should be arbitrarily small.

 The discrete setting --- the case where the correlated sources take values in a finite alphabet --- was studied by Csisz\'{a}r and Narayan~\cite{Csiszar_multiterminal}.
They gave a scheme for computing a secret key in  this setting and found the secret key capacity, i.e., the maximum achievable secret key rate.
This was later generalized by Nitinawarat and Narayan~\cite{Nitinawarat12} to the case where the terminals possess correlated Gaussian sources.

In a practical setting, we can assume that the random sources are obtained by observing some natural parameters, e.g., temperature in a field.
In other words, the underlying source is continuous. However, for the purposes of storage and computation, these sources must be quantized, and only the quantized source can be used for secret key generation. 
If each terminal uses a scalar quantizer, then we get the discrete source model studied in~\cite{Csiszar_multiterminal}.
However, we could do better and instead use a vector quantizer to obtain a higher secret key rate. Nitinawarat and Narayan~\cite{Nitinawarat12} found the secret key capacity for correlated Gaussian sources in such a setting. However, to approach capacity, the quantization rate and the rate of public communication at each terminal must approach infinity. In practice, it is reasonable to have a constraint on the quantization rate at each terminal.
The terminals can only use the quantized source for secret key generation. 
Nitinawarat and Narayan~\cite{Nitinawarat12} studied a two-terminal version of this problem, where a quantization rate constraint was imposed on only one of the terminals. They gave a nested lattice coding scheme and showed that it was optimal, i.e., no other scheme can give a higher secret key rate. 
In related work, Watanabe and Oohama~\cite{Watanabe11} characterized the maximum secret key rate achievable under a constraint on the 
rate of public communication in the two-terminal setting. More recently, Ling et al.~\cite{Ling13} gave a lattice coding scheme for the public communication-constrained problem and were able to achieve a secret key rate  within $1/2$ nats of the maximum in~\cite{Watanabe11}. 

We consider a multiterminal generalization of the two-terminal version studied by~\cite{Nitinawarat12} where quantization rate constraints are imposed on each of the terminals.
Terminal $i$ has access to  $N$ iid copies of a Gaussian source $X_i(1),X_i(2),\ldots,X_i(N)$. The sources are correlated across the terminals.
We assume that the joint distribution of the sources has a \emph{Markov tree} structure~\cite[Example 7]{Csiszar_multiterminal}, which is a generalization of a Markov chain.
Let us define this formally.
Suppose that $T=(V,E)$ is a tree and $\{X_i:i\in V\}$ is a collection of random variables indexed by the vertices.
Consider any two disjoint subsets $\mathcal{I}$ and $\mathcal{J}$ of $V$. Let $v$ be any vertex such that removal of $v$ from $T$ disconnects $\mathcal{I}$ from $\mathcal{J}$ (Equivalently, for every $i\in \mathcal{I}$ and $j\in \mathcal{J}$, the path connecting $i$ and $j$ passes through $v$). For every such $\mathcal{I},\mathcal{J},\tv$, if $\{X_i:i\in\mathcal{I} \}$ and $\{ X_j:j\in \mathcal{J} \}$ are conditionally independent given $X_v$, then we say that $\{X_i:i\in T\}$ form a Markov chain on $T$. 
Alternatively, we say that $\{X_i:i\in V\}$ is a Markov tree source.


The contributions of this paper are the following. We study the problem of secret key generation in a Gaussian Markov tree source model with individual quantization rate constraints imposed at each terminal.
We give a nested lattice-based scheme and find the achievable secret key rate. For certain classes of Markov trees, particularly homogeneous Markov trees\footnote{We say that a Markov tree is homogeneous if $I(X_{\tu};X_{\tv})$ is the same for all edges $(\tu,\tv)$}, we show that our scheme achieves the secret key capacity as the quantization rates go to infinity. However, we also give examples where our scheme does not achieve the key capacity. A salient feature of our scheme is that the overall computational complexity required for quantization and key generation is polynomial in the number of samples $N$. It is also interesting to note that unlike the general schemes in~\cite{Csiszar_multiterminal,Nitinawarat12}, we give a scheme where at least one terminal remains silent (does not participate in public communication), and omniscience is not attained.

\section{Notation and Definitions}\label{sec:notation}
If $\mathcal{I}$ is an index set and $\{A_i: i\in \mathcal{I} \}$ is a class of sets indexed by $\mathcal{I}$, then their Cartesian product is denoted by $\bigtimes_{i\in \mathcal{I}}A_i$. Given two sequences indexed by $n\in \mathbb{N}$, $f(n)$ and $g(n)$, we say that $f(n)=O(g(n))$ if there exists a constant $c$ such that $f(n)\leq cg(n)$ for all sufficiently large $n$. Furthermore, $f(n)=o_n(1)$ if $f(n)\to 0$ as $n\to\infty$.


Let $G=(V,E)$ be a graph. The distance between two vertices $\tu$ and $\tv$ in $G$ is the length of the shortest path between $\tu$ and $\tv$.  
Given a rooted tree $T=(V,E)$ with root $\mathtt{r}(T)$ we say that a vertex $\tu$ is the parent of $\tv\neq \mathtt{r}(T)$, denoted $\tu=\parent(\tv)$, if $\tu$ lies in the shortest path from $\mathtt{r}(T)$ to $\tv$ and the distance between $\tu$ and $\tv$ is $1$. Furthermore, for every $\tv\in V$, we define $N_{T}(\tv)$ to be the set of all neighbours of $\tv$ in $T$.
\section{Secret Key Generation from Correlated Gaussian Sources}

\subsection{The Problem}\label{sec:prob_definition}
We now formally define the problem. We consider a multiterminal Gaussian source model~\cite{Nitinawarat12}, which is described as follows. There are $m$ terminals, each having access to $N$ independent and identically distributed (iid) copies of a correlated Gaussian source, i.e., the $l$th terminal observes $X_l(1),X_l(2),\ldots,X_{l}(N)$ which are iid. Without loss of generality, we can assume that $X_l(i)$ has mean zero and variance $1$. We can always subtract the mean and divide by the variance to ensure that this is indeed the case.  The joint distribution of $\{X_{l}(i):1\leq l\leq m \}$ can be described by their covariance matrix $\Phi$. 

Specifically, we assume that the sources form a Markov tree, defined in Sec.~\ref{sec:introduction}. Let $T=(V,E)$ be a tree having $|V|=m$ vertices, which defines the conditional independence structure of the sources.  For $\tu, \tv\in V$, let us define $\rhouv:=\bE[X_{\tu}X_{\tv}]$.

We can therefore write
\[
	X_{\tu} = \rhouv X_{\tv} + \sqrt{1-\rhouv^2}\; Z_{\tu\tv}
\]
where $Z_{\tu\tv}$ is a zero-mean, unit-variance Gaussian random variable which is independent of $X_{\tv}$. Similarly,
\[
	X_{\tv} = \rhouv X_{\tu} + \sqrt{1-\rhouv^2}\; Z_{\tv\tu}
\]
where $Z_{\tv\tu}$ is also a zero-mean, unit-variance Gaussian random variable which is independent of $X_{\tu}$ (and different from $Z_{\tu\tv}$). 

Our objective is to generate a secret key using public communication.
For $\tv\in V$, let $\XNv:=(X_{\tv}(1),X_{\tv}(2),\ldots,X_{\tv}(N))$
denote the $N$ iid copies of $X_{\tv}$ available at terminal $\tv$.  Each terminal uses a vector quantizer $Q_{\tv}:\R^N\to \mathcal{X}_{\tv}$ of rate $\Rqv:=\frac{1}{N}\log_2|\mathcal{X}_{\tv}|$. Terminal $\tv$ transmits $\FNv\in \mathcal{F}^{(N)}_{\tv}$ --- which is a (possibly randomized) function of $Q_{\tv}(\XNv)$ --- across a noiseless public channel that an eavesdropper may have access to\footnote{In this work, we only consider noninteractive communication, i.e., the public communication is only a function of the source and not of the prior communication.}. Using the public communication and their respective observations of the quantized random variables, $Q_{\tv}(\XNv)$, the terminals must generate a secret key $\KN\in \mathcal{K}^{(N)}$ which is concealed from the eavesdropper. Let $\mathcal{F}_G:=\bigtimes_{\tv\in V}\mathcal{F}_{\tv}^{(N)}$. 

Fix any $\epsilon>0$. We say that $\KN$ is an $\epsilon$-secret key ($\epsilon$-SK) if there exist functions $f_{\tv}:(\mathcal{X}_{\tv},\mathcal{F}_{G})\to \mathcal{K}^{(N)}$ such that:
$$\Pr\left[f_{\tv}(Q_{\tv}(\XNv),\{\mathbf{F}_{\tu}^{(N)}:\tu\in V\})\neq \KN\right]<\epsilon,$$
$$\log_2|\mathcal{K}^{(N)}|-H(\KN)<\epsilon,$$ and
 $$I\left(\{\FNv:\tv\in V\};\KN\right)<\epsilon.$$

We say that $\Rkey$ is an achievable secret key rate if for every $\epsilon>0$, there exist quantizers $\{Q_{\tv}\}$, a scheme for public communication, $\{ \FNv \}$, and a secret key $\KN$, such that for all sufficiently large $N$, $\KN$ is an $\epsilon$-SK, and $\frac{1}{N}\log_{2}|\mathcal{K}^{(N)}|\geq \Rkey - \epsilon$.

\begin{figure}
\begin{center}
\includegraphics[width=7cm]{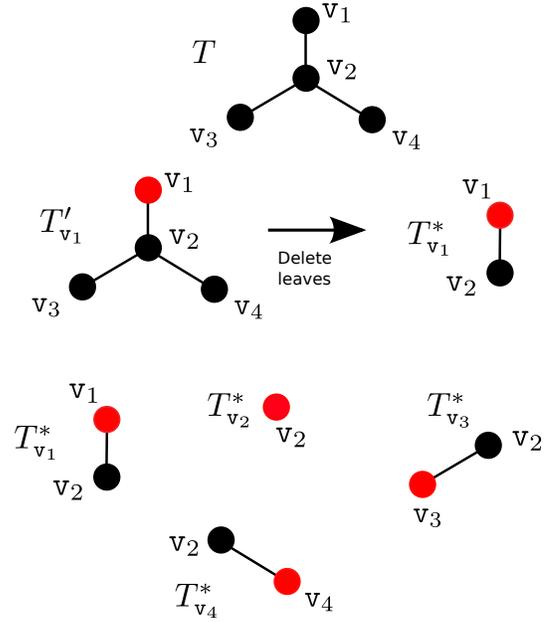}
\end{center}
\caption{Illustration of $\cTst$ for a tree having four vertices.}
\label{fig:Gstar}
\end{figure}

Consider the following procedure to obtain a class of rooted subtrees of $T$:
\begin{itemize}
\item Identify a vertex $\tv$ in $V$ as the root. The tree $T$ with $\tv$ as the root is a rooted tree. Call this $T_{\tv}'$.
\item Delete all the leaves of $T_{\tv}'$. Call the resulting rooted subtree $\Tst_{\tv}$.
\end{itemize}
Let $\cTst:=\{ \Tst_{\tv}: \tv\in V \}$ denote the set of all rooted subtrees of $T$ obtained in the above manner. Fig.~\ref{fig:Gstar} illustrates this for a tree having four vertices.
 Note that there are $|V|$ trees in $\cTst$, one corresponding to each vertex. 
 For any such rooted subtree $\Tst=(V^*,E^*)$ in $\cTst$, let $\rTst$ denote the root of $\Tst$.
  We will see later that it is only the terminals that correspond to $\Tst$ that
participate in the public communication while the other terminals remain silent.
For any $\tv\in V^*$, let $N_T(\tv)$ denote the set of all neighbours of $\tv$ in $T$ (\emph{not} $\Tst$).
Recall that each terminal $\tv$ operates under a quantization rate constraint of $\Rqv$.
For every $\Tst=(V^*,E^*)$, us define 
\begin{align}
\mathcal{R}_{\mathrm{ent}} &= R_{q}^{(\rTst)} &\notag \\
                         & \quad +\sum_{\tu\in V^*\backslash \rTst}\frac{1}{2}\log_2\left((e^{2\Rqu}-1)(1-\rho_{\tu, \parent(\tu)}^2)+1\right)\label{eq:Rent}
\end{align}
and 
\begin{multline}
\mathcal{R}_{\mathrm{com}}= \sum_{\tv\in V^*}\max_{\tu \in N_{T}(\tv)} \frac{1}{2}\log_2\Bigg((e^{2\Rqv}-1)(1-\rho_{\tu \tv}^2)\\+1+\frac{\rho_{\tu \tv}^2e^{2\Rqv}}{e^{2\Rqu}-1} \Bigg).\label{eq:Rcom}
\end{multline}
We will show that the joint entropy of the quantized sources is at least $\mathcal{R}_{\mathrm{ent}}$ and the sum rate of public communication is  at most $\mathcal{R}_{\mathrm{com}}$ in our scheme. Also, the public communication that achieves $\mathcal{R}_{\mathrm{com}}$ requires only the terminals in $T^*$
to participate in the communication; the terminals in $V\backslash V^*$ are silent.
Let us also define 
\begin{equation}
\alpha:= \frac{\max_{\tu\in V^*}\Rqu}{\min_{\tv\in V^*}\Rqv}.
\end{equation}
 Our aim is to prove the following result

\begin{theorem}
For a fixed quantization rate constraint $\{ \Rqv:\tv\in V \}$, a secret key rate of 
\begin{align}
\Rkey &= \max_{\Tst\in\cTst}\Bigg\{  \mathcal{R}_{\mathrm{ent}}-\mathcal{R}_{\mathrm{com}} \Bigg\} &\label{eq:ach_SKrates}
\end{align}
is achievable using a nested lattice coding scheme whose computational complexity grows as $O(N^{\alpha+1})$.
\label{thm:main}
\end{theorem}
Note that if all terminals have identical quantization rate constraints, then the complexity is $O(N^2)$.
Sec.~\ref{sec:scheme} describes the scheme and contains the proof of the above theorem.

We now  discuss some of the implications of the result.
Letting the quantization rates $\Rqu$ in (\ref{eq:ach_SKrates}) go to infinity, i.e., as $\Rqv\to\infty$ for all $\tv$, we get that
\begin{corollary}
In the fine quantization limit, a  secret key rate of 
\begin{align}
\Rkey &= \max_{\Tst\in\cTst} \Bigg\{ \min_{\tv\in N_{T}(\rTst)}\frac{1}{2}\log_2\left( \frac{1}{1-\rho_{\rTst \tv}^2}\right)& \notag\\
                  &  \qquad  + \sum_{\tu\in V^*\backslash \rTst}\min_{\tv \in N_{T}(\tu)} \frac{1}{2}\log_2\left( \frac{1-\rho_{\tu,\parent(\tu)}^2}{1-\rho_{\tu,\tv}^2} \right) \Bigg\} \label{eq:ach_SKrates_fine}
\end{align}
is achievable.
\label{cor:ach_rate_finequantization}
\end{corollary}
If there are no constraints on the quantization rates, then from~\cite[Theorem 3.1]{Nitinawarat12} and~\cite[Example 7]{Csiszar_multiterminal}, we know that
the maximum achievable secret key rate is 
\begin{equation}
C_{\mathrm{key}}^{(\infty)} = \min_{(\tu,\tv)\in E}\frac{1}{2}\log_2\left(\frac{1}{1-\rhouv^2}\right).
\label{eq:Ckey}
\end{equation}

\section{Remarks on the Achievable Secret Key Rate}
\subsection{The Two-User Case}
Consider the two-user case with  terminals $\tu$ and $\tv$.
Let us define 
\[
 \mathcal{R}(\tu,\tv):= \frac{1}{2}\log_2\left(\frac{e^{2\Rqu}}{(e^{2\Rqu}-1)(1-\rho_{\tu \tv}^2)+1+\frac{\rho_{\tu \tv}^2e^{2\Rqu}}{e^{2\Rqv}-1}} \right)
\]
As we will see later, the above SK rate is achieved with $\tu$ participating in the public communication and $\tv$ remaining silent.
The achievable secret key rate, (\ref{eq:ach_SKrates}), is equal to $\max\{\mathcal{R}(\tu,\tv),\mathcal{R}(\tv,\tu)\}$.
A simple calculation reveals that
\begin{multline}
 e^{-2\mathcal{R}(\tu,\tv)}-e^{-2\mathcal{R}(\tv,\tu)}\\=\rhouv^2\left( \frac{1}{e^{2\Rqv}(e^{2\Rqv}-1)}-\frac{1}{e^{2\Rqu}(e^{2\Rqu}-1)} \right)
\end{multline}
Hence, if $\Rqv>\Rqu$, then $\mathcal{R}(\tu,\tv)>\mathcal{R}(\tv,\tu)$. This means that in order to obtain a higher secret key rate using our scheme, the terminal with the lower quantization rate must communicate, while the other must remain silent.

\begin{figure}
\includegraphics[width=9cm]{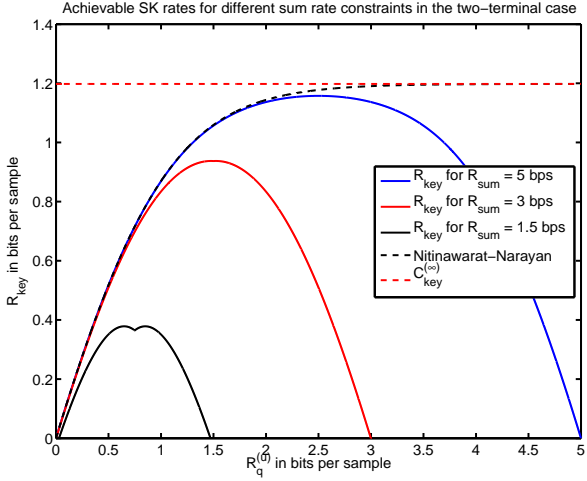}
\caption{Plot of achievable secret key rates under a sum rate constraint for two terminals.}
\label{fig:achrates_twouser_sumrate}
\end{figure}

If we let $\Rqv$ in $\mathcal{R}(\tu,\tv)$ go to infinity, then we get the rate $R_{\mathrm{NN}}$ achieved in~\cite{Nitinawarat12}, which was shown to be optimal when we only restrict the quantization rate of one terminal.
\[
	R_{\mathrm{NN}} = \frac{1}{2}\log_2\left(\frac{e^{2\Rqu}}{(e^{2\Rqu}-1)(1-\rho_{\tu \tv}^2)+1} \right).
\]
Fig.~\ref{fig:achrates_twouser_sumrate} illustrates the behaviour of the achievable rate for different sum-rate constraints ($\Rqu+\Rqv=R$).
The rate achieved by the scheme of Nitinawarat and Narayan~\cite{Nitinawarat12}, $R_{\mathrm{NN}}$, is also shown.

\subsection{Optimality of $\Rkey$ in the Fine Quantization Limit}
We present a class of examples where $\Rkey$ is equal to the secret key capacity $C_{\mathrm{key}}^{(\infty)}$ in the fine quantization limit.
One such example is the class of \emph{homogeneous} Markov trees, where $\rhouv=\rho$ for all edges $(\tu,\tv)$. In this case, $$\min_{\tv \in N_{T}(\tu)} \frac{1}{2}\log_2\left( \frac{1-\rho_{\tu,\parent(\tu)}^2}{1-\rho_{\tu,\tv}^2} \right) = 0,$$
and hence, by Corollary~\ref{cor:ach_rate_finequantization}, $$\Rkey =\frac{1}{2}\log_2\left( \frac{1}{1-\rho^2} \right)=C_{\mathrm{key}}^{(\infty)}.$$

This property holds for a wider class of examples. Consider the case where $T$ has a rooted subtree $\Tst$ such that for every $\tu\in V^*$, $\arg\min_{\tv\in N_T(\tu)}\rhouv=\parent(\tu)$. Once again, we have
$$\min_{\tv \in N_{T}(\tu)} \frac{1}{2}\log_2\left( \frac{1-\rho_{\tu,\parent(\tu)}^2}{1-\rho_{\tu,\tv}^2} \right) = 0.$$
Moreover, the edge $(\tu,\tv)\in E$ with the minimizing $\rhouv$ (and therefore, the minimizing mutual information) is incident on $\rTst$. 
Hence, $\Rkey = C_{\mathrm{key}}^{(\infty)}$.
\subsection{Suboptimality of $\Rkey$ in the Fine Quantization Limit}
\begin{figure}
\includegraphics[width=7.5cm]{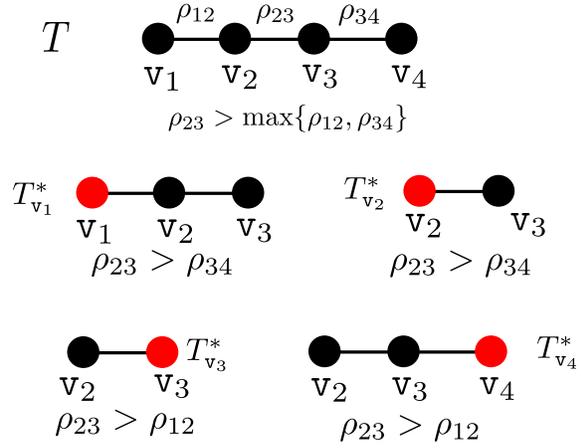}
\caption{An example where our scheme is suboptimal.}
\label{fig:suboptimaleg}
\end{figure}
We can give several examples for which $\Rkey$ in the fine quantization limit is strictly less than $C_{\mathrm{key}}^{(\infty)}$. Note that
\[
\min_{\tv \in N_{T}(\tu)} \frac{1}{2}\log_2\left( \frac{1-\rho_{\tu,\parent(\tu)}^2}{1-\rho_{\tu,\tv}^2} \right)\leq 0,
\]
and  if these terms are nonzero for every $\Tst\in \cTst$, then the scheme is suboptimal. As a specific example, consider the Markov chain of Fig.~\ref{fig:suboptimaleg},
where $\rho_{23}>\max\{\rho_{12},\rho_{34}\}$. Let us further assume that $\rho_{12}=\rho_{34}$. The secret key capacity is
\[
 C_{\mathrm{key}}^{(\infty)} = \frac{1}{2}\log_2\frac{1}{1-\rho_{12}^2}.
\] 
Irrespective of which $\Tst\in \cTst$ we choose (see Fig.~\ref{fig:suboptimaleg}), we have 
\[
\min_{\tv\in N_{T}(\rTst)}\frac{1}{2}\log_2\left( \frac{1}{1-\rho_{\rTst \tv}^2}\right) = C_{\mathrm{key}}^{(\infty)}.
\]
Furthermore, the second term in (\ref{eq:ach_SKrates_fine}) is negative for every $\Tst$. This is because every $\Tst$ has some $\tu\neq \rTst$ for which 
$\arg\min_{\tv\in V^*}\rhouv\neq\parent(\tv)$.
\section{The Secret Key Generation Scheme}\label{sec:scheme}
We now describe the lattice coding scheme that achieves the promised secret key rate.
Our scheme is very similar to the scheme given by Nitinawarat and Narayan~\cite{Nitinawarat12} for the 
two-terminal case. 

We use a block encoding scheme just like the one in~\cite{Nitinawarat12}.
Recall that each terminal $\tv$ has a quantization rate constraint of $\Rqv$.
The total blocklength $N$ is partitioned into $\Nout$ blocks of $n$ samples each, i.e., $N=n\Nout$, where $\Nout = \min_{\tv}2^{n\Rqv}-1$.
The secret key generation scheme comprises  two phases: an information reconciliation phase,
and a privacy amplification phase. The reconciliation phase is divided into two subphases: a lattice coding-based analog phase,
which is followed by a Reed-Solomon coding-based digital phase. The privacy amplification phase employs a linear mapping to generate the secret key
from the reconciled information. This uses the results of Nitinawarat and Narayan~\cite{Nitinawarat12}. The digital phase is also inspired by the 
concatenated coding scheme used in~\cite{Vatedka_concat} in the context of channel coding  for Gaussian channels.
 
Let us briefly outline the protocol for secret key generation. Each terminal $\tv$ uses a chain of nested lattices $(\Lv,\Lva,\Lvb)$ in $\R^n$, where $\Lvb\subset\Lva\subset \Lv$. The Gaussian input $\xv$ at terminal $\tv$ is processed blockwise, with $n$ samples collected to form a block. Suppose $\xv=(\xv^{(1)},\ldots,\xv^{(\Nout)})$ where $\xvi$ denotes the $i$th block of length $n$. Each terminal $\tv$ also generates random dithers $\{\dvi: 1\leq i\leq \Nout \}$, which are all uniformly distributed over the fundamental Voronoi region of $\Lv$, and independent of each other. These are assumed to be known to all terminals.\footnote{In principle, the random dither is not required. Similar to~\cite{Nazer11}, we can show that there exist fixed dithers for which all our results hold. One could avoid the use of dithers by employing the techniques in~\cite{Ling13}, but we do not take that approach here.} The protocol for secret key generation is as follows. 
\begin{itemize}
\item \emph{Quantization:} Terminal $\tv\in V$ computes
\[
	\yvi = \big[ Q_{\Lv}(\xvi+\dvi)-\dvi \big]\bmod\Lvb.
\]
\item \emph{Information reconciliation: Analog phase:} Let $T^*=(V^*,E^*)$ be the rooted subtree which achieves the maximum in (\ref{eq:ach_SKrates}). Terminal $\tv\in V^*$ broadcasts
\[
	\wvi = [\yvi]\bmod\Lva.
\]
across the public channel. Terminal $\tu$ has access to $\yui$ and for every $\tv\in N_{T^*}(\tu)$, it estimates
\[
	\widehat{\y}_{\tv}^{(i)} = \wvi+ Q_{\Lva}\left( \rhouv \yui - \wvi \right).
\]
Having estimated $\yvi$ for all neighbours $\tv$, it estimates $\yvi$ for all $\tv$ which are at a distance $2$ from $\tu$, and so on, till it has estimated $\{ \yvi:\tv\in V^*, 1\leq i\leq \Nout \}$.
\item \emph{Information reconciliation: Digital phase:} To ensure that all $\Nout$ blocks can be recovered at all terminals with an arbitrarily low probability of error, we use a Slepian-Wolf scheme using Reed-Solomon codes. Each terminal uses an $(\Nout,K_{\mathrm{out}})$ Reed-Solomon code over $\Fpkv$, where the parameters $K_{\mathrm{out}}$ and $p^{k_{\tv}}$ will be specified later. The syndrome corresponding to\footnote{We show that there is a bijection between $\Lv/\Lvb$ and $\Fpkv$.} $(\y_{\tv}^{(1)},\ldots,\y_{\tv}^{(\Nout)})$ in the code is publicly communicated by terminal $\tv$. We show that this can be used by the other
terminals to estimate all the $\yvi$s with a probability of error that decays exponentially in $N$.
\item \emph{Key generation:} We use the result~\cite[Lemma 4.5]{Nitinawarat12} that there exists a linear transformation of the source symbols (viewed as elements of a certain finite field) that can act as the secret key. Since all terminals can estimate $\{\yvi:\tv\in V^*,1\leq i\leq \Nout  \}$ reliably, they can all compute the secret key with an arbitrarily low probability of error.
\end{itemize}

Before we go into the details of each step, we describe some specifics of the coding scheme.
We want the nested lattices that form the main component of our protocol to satisfy certain ``goodness'' properties. We begin by describing the features that the lattices must possess. 

\subsection{Nested Lattices}\label{sec:nestedlattice}
Some basic definitions and relevant results on lattices have been outlined in Appendix~\ref{sec:latticeconcepts}.
Given a lattice $\L$, we let $\cV(\L)$ be the fundamental Voronoi region of $\L$, and $\sigma^2(\L)$ denotes the second moment per dimension of $\L$.
Furthermore, we define $\vol(\L):=\vol(\cV(\L))$.
 

Each terminal $\tv$ uses a chain of $n$-dimensional nested lattices $(\Lv,\Lva,\Lvb)$, with $\Lvb\subset\Lva\subset \Lv$.
These are all Construction-A lattices~\cite{Erez04,Erez05} obtained from linear codes of blocklength $n$ over $\Fp$, with $p$ chosen large enough to ensure that these lattices satisfy the required goodness properties.
Furthermore, $\Lva$ and $\Lvb$ are obtained from subcodes of linear codes that generate $\Lv$.
Fix any $\delta>0$.
The lattices are chosen so that
\begin{equation}
\frac{1}{n}\log_2|\Lv\cap\cV(\Lvb)|=\frac{1}{n}\log_2\frac{\text{vol}(\Lvb)}{\text{vol}(\Lv)}=\frac{k_{\tv}}{n}\log_2 p=\Rqv,
\label{eq:quantizationrateconstraint}
\end{equation}
\begin{equation}
\frac{\big(\vol(\Lvb)\big)^{2/n}}{2\pi e} = \big( 1+ \sigma^2(\Lv) \big)(1+\delta),
\label{eq:vol_Lvb_constraint}
\end{equation}
and
\begin{align}
\frac{\big(\vol(\Lva)\big)^{2/n}}{2\pi e} &= \max_{\tu\in N_{T}(\tv)} \Big( {1-\rhouv^2} +\sigma^2(\Lv)+\rhouv^2\sigma^2(\Lu)  \Big)&\notag\\
                                           &\hspace{3cm} \times(1+\delta).
\label{eq:vol_Lva_constraint}
\end{align}
Furthermore, these lattices satisfy the following ``goodness'' properties~\cite{Erez05}:
\begin{itemize}
\item $\Lv$ is good for covering.
\item $\Lva$ and $\Lvb$ are good for AWGN channel coding.
\end{itemize}

\subsection{Quantization}\label{sec:quantization}
Terminal $\tv$ observes $N$ samples $\xv = (x_{\tv}(1),\ldots,x_{\tv}(N))$.
As mentioned earlier, the quantizer operates on blocks of $n$ samples each, and there are $\Nout$ such blocks.
We can write $\x = (\xu^{(1)},\xu^{(2)},\ldots,\xu^{(\Nout)})$, where
$\xu^{(j)}\in \R^n$ is given by $\xu^{(j)}=(x_{\tu}((j-1)n+1),\ldots,x_{\tu}(jn))$.

Terminal $\tv$ also generates $\Nout$ dither  vectors $\dv^{(1)},\dv^{(2)},\ldots,\dv^{(\Nout)}$, which are all uniformly distributed over $\cV(\Lv)$, and independent of each other and of everything else. These dither vectors are assumed to be known to all the terminals, and to the eavesdropper. 

For $1\leq i\leq \Nout$ and $\tv\in V$, let
\begin{equation}
\yvi = [Q_{\Lv}(\xvi+\dvi)-\dvi]\bmod\Lvb
\label{eq:quantized_output}
\end{equation}
denote the output of the lattice quantizer at terminal $\tv$. The terminals can only use $\yv:=(\yv^{(1)},\yv^{(2)},\ldots,\yv^{(\Nout)})$ for the secret key generation protocol.
From (\ref{eq:quantizationrateconstraint}) and (\ref{eq:vol_Lvb_constraint}), we can see that the quantization rates satisfy
\begin{equation}
\Rqv = \frac{1}{2}\log_2\left( 1+\frac{1}{\sigma^2(\Lv)} \right)+\log_2(1+\delta)+o_n(1).
\label{eq:quantizationrate}
\end{equation}

\subsection{Information Reconciliation: The Analog Phase}\label{sec:analog_stage}
Let $\Tst=(V^*,E^*)$ denote the rooted tree in $\cTst$ which achieves the maximum in (\ref{eq:ach_SKrates}).
The terminals in $V^*$ are the only ones that communicate across the public channel. Terminal $\tv\in V^*$ broadcasts
\begin{align}
\wvi &:= [\yvi]\bmod\Lva &\notag\\
	& = [Q_{\Lv}(\xvi+\dvi)]\bmod\Lva &\label{eq:comm_rv}
\end{align}
for $1\leq i\leq \Nout$, across the public channel.
Prior to the analog phase, terminal $\tu\in V$ only has access to $\yu=(\yu^{(1)},\yu^{(2)},\ldots,\yu^{(\Nout)})$.
At the end of the information reconciliation phase, every terminal $\tu$ will be able to recover $\{ \yv:\tv\in V^* \}$ with low probability of error. The analog phase ensures that every $\yvi$ can be individually recovered with low probability of error. The digital phase guarantees that the entire block $\yv$ can be recovered reliably.

Now consider any $\tv\in V^*$ and $\tu\in N_{T}(\tv)$ (not necessarily in $V^*$).
Suppose that some terminal $\mathtt{u}'$ (not necessarily $\tu$) has a reliable estimate of $\yui$. From $\yui$ and $\wvi$, terminal $\tu'$ can estimate $\yvi$ as follows:
\begin{equation}
 \hyvi =\; \wvi \; +\;  Q_{\Lva}\left(\rhouv\yui-\wvi\right).
\end{equation} 
The following proposition is proved in Appendix~\ref{sec:proof_prop_analog}.
\begin{proposition}
Fix a $\delta>0$. For every $1\leq i\leq\Nout$, we have
\begin{equation}
\mathbb{E}_{\yui}\Pr[\hyvi\neq \yvi] \leq e^{-nE_{\tu\tv}(\delta)}
\end{equation}
where $E_{\tu\tv}$ is a quantity which is positive for all positive $\delta$ and all sufficiently large $n$, as long as 
\begin{align}
\frac{\big(\vol(\Lva)\big)^{2/n}}{2\pi e} &> \max_{\tu\in N_{T}(\tv)} \Big( {1-\rhouv^2} +\sigma^2(\Lv)+\rhouv^2\sigma^2(\Lu)  \Big)&\notag\\
                                     & \hspace{3cm} \times (1+\delta), &\label{eq:volLva_constraint}
\end{align}
\begin{equation}
\frac{\big(\vol(\Lvb)\big)^{2/n}}{2\pi e} > (1+\sigma^2(\Lv))(1+\delta),
\label{eq:volLvb_constraint}
\end{equation}
and
\begin{equation}
\frac{\big(\vol(\Lub)\big)^{2/n}}{2\pi e} > (1+\sigma^2(\Lu))(1+\delta).
\label{eq:volLub_constraint}
\end{equation}
\label{prop:Pe_analog}
\end{proposition}
Since terminal $\tu$ has $\yui$, it can (with high probability) recover the corresponding quantized sources of its neighbours. 
Assuming that these have been recovered correctly, it can then estimate the quantized sources of all terminals at distance two from $\tu$, and so on, till all $\yvi$ for $\tv$ in $V^*$ have been recovered.
Using the union bound, we can say that the probability that terminal $\tu$ correctly recovers $\{ \yvi: \tv\in V^* \}$ is at least $1-\sum_{\tu\in V^*}\max_{\tv\in N_{T}(\tu)}e^{-nE_{\tu\tv}(\delta)}$.

For all terminals to be able to agree upon the key, we must ensure that every terminal can recover all blocks $\{ \yv:\tv\in V^* \}$ with low probability of error. Since $\Nout$ is exponential in $n$, the analog phase does not immediately guarantee this.
For that, we use the digital phase.

\subsection{Information Reconciliation: The Digital Phase}\label{sec:digitalstage}
Observe that $\yvi\in \Lv\cap\cV(\Lvb)$, where both $\Lv$ and $\Lvb$ are Construction-A lattices obtained by 
linear codes over $\Fp$. As a result, $|\Lv\cap\cV(\Lvb)|$ is always an integer power of $p$~\cite{Erez05}.
Let 
\[
|\Lv\cap\cV(\Lvb)|=p^{k_{\tv}}.
\]
Then, there exists an (set) isomorphism $\varphi_{\tv}$ from $\Lv\cap\cV(\Lvb)$ to $\Fpkv$.
For every $\tv\in V^*$ and $i\in \{1,2,\ldots,\Nout\}$, let $\fyvi=\varphi_{\tv}(\yvi)$.
Similarly, let $\hfyvi = \varphi_{\tv}(\hyvi)$. 

The key component of the digital phase is a Reed-Solomon code over $\Fpkv$.
In~\cite{Nitinawarat12}, a Slepian-Wolf scheme with random linear codes was used for the digital phase.
Using a Reed-Solomon code, we can ensure that the overall computational complexity (including all the phases of the protocol) is polynomial in $N$. 

For every $\tv$, let $\mathcal{C}_{\tv}$ be a Reed-Solomon code of blocklength $\Nout$ and dimension
\begin{equation}
K_{\mathrm{out}} = \Nout (1-2\delta).
\end{equation}
Let $\mathsf{y}_{\tv}^{\Nout}=(y_{\tv}^{(1)},y_{\tv}^{(2)},\ldots , y_{\tv}^{(\Nout)})$ and
$\widehat{\mathsf{y}}_{\tv}^{\Nout}=(\widehat{y}_{\tv}^{(1)},\widehat{y}_{\tv}^{(2)},\ldots , \widehat{y}_{\tv}^{(\Nout)})$. 
We can write
\[
	\widehat{\mathsf{y}}_{\tv}^{\Nout} = \mathsf{y}_{\tv}^{\Nout} + \mathsf{e}_{\tv}^{\Nout},
\]
where $\mathsf{e}_{\tv}^{\Nout}=(e_{\tv}^{(1)},e_{\tv}^{(2)},\ldots,e_{\tv}^{(\Nout)})$ is the error vector, and from the previous section, we have
$$\Pr[e_{\tv}^{(i)}\neq 0]\leq \sum_{\tv\in V^*}\max_{\tu\in N_{T}(\tv)}e^{-nE_{\tu\tv}(\delta)}\leq \delta$$
 for all sufficiently large $n$.
Every $\mathsf{y}_{\tv}^{\Nout}$ can be written uniquely as 
\begin{equation}
	\mathsf{y}_{\tv}^{\Nout} = \mathsf{c}_{\tv}^{\Nout} + \mathsf{s}_{\tv}^{\Nout}
\label{eq:yvout}
\end{equation}
where $\mathsf{c}_{\tv}^{\Nout}\in \mathcal{C}_{\tv}$, and $\mathsf{s}_{\tv}^{\Nout}$ is a minimum Hamming weight representative of the coset to which $\mathsf{y}_{\tv}^{\Nout}$ belongs in $\Fpkv^{\Nout}/\mathcal{C}_{\tv}$. Terminal $\tv$ broadcasts $\mathsf{s}_{\tv}^{\Nout}$ across the public channel. This requires a rate of public communication of at most
\begin{equation}
\frac{1}{N}\log_2|\Fpkv^{\Nout}/\mathcal{C}_{\tv}| = \frac{2\Nout \delta}{N}\log_2 (p^{k_{\tv}})= 2\delta \Rqv.
\label{eq:digital_leakage}
\end{equation}
 From $\mathsf{s}_{\tv}^{\Nout}$ and $\widehat{\mathsf{y}}_{\tv}^{\Nout}$, terminal $\tu$ can compute
\[
\widehat{\mathsf{c}}_{\tv}^{\Nout}= \widehat{\mathsf{y}}_{\tv}^{\Nout}-\mathsf{s}_{\tv}^{\Nout}= \mathsf{c}_{\tv}^{\Nout} + \mathsf{e}_{\tv}^{\Nout}.
\]

For sufficiently large $n$ the probability that $\yvi$ is estimated incorrectly is less than $\delta$, and terminal $\tu$ can recover $\mathsf{c}_{\tv}^{\Nout}$ with high probability using the decoder for the Reed-Solomon code. 

\begin{proposition}[Theorem 2, \cite{Vatedka_concat}]
 The probability that the Reed-Solomon decoder incorrectly decodes $\mathsf{c}_{\tv}^{\Nout}$ from $\widehat{\mathsf{c}}_{\tv}^{\Nout}$ decays exponentially in $N$.
 \label{prop:Pe_digital}
\end{proposition}
Having recovered $\mathsf{c}_{\tv}^{\Nout}$ reliably, the terminals can obtain $\mathsf{y}_{\tv}^{\Nout}$ using (\ref{eq:yvout}).
Therefore, at the end of the digital phase, all terminals can recover $\{ \yv:\tv\in V^* \}$ with a probability of error that decays exponentially in $N$.

\subsection{Secret Key Generation}\label{sec:key_generation}
Let $k:=\sum_{\tv\in V^*}k_{\tv}$. There exists a (set) bijection $\phi$ from $\bigtimes_{\tv\in V^*}\Fpkv$ to $\mathbb{F}_{p^k}$.
Let $y^{(i)}=\phi(\yvi:\tv\in V^*)$. 
We use the following result by Nitinawarat and Narayan~\cite{Nitinawarat12}, which says that there exists a linear function of the sources that can act as the secret key.

\begin{lemma}[Lemma 4.5, \cite{Nitinawarat12}]
 Let $Y$ be a random variable in a Galois field  $\mathbb{F}_q$ and $D$ be an $\R^n$-valued random variable jointly distributed with $Y$.
 Consider $\Nout$ iid repetitions of $(Y,D)$, namely $(Y^{\Nout},D^{\Nout}) = ((Y_1 , D_1 ) , \ldots , (Y_{\Nout} , D_{\Nout} ))$. 
 
 Let
 $B = B^{(\Nout)} \in \mathcal{B}^{(\Nout)}$ be a finite-valued rv with a given joint distribution with
 $(Y^{\Nout},D^{\Nout})$. 
 
 Then, for every $\delta>0$ and every $$R < H(Y|D ) -\frac{1}{\Nout}\log |\mathcal{B}^{(\Nout)}|-2\delta, $$
there exists a
 $ \lfloor \frac{\Nout R}{\log q} \rfloor\times \Nout$ matrix $L$ with $\mathbb{F}_q$ -valued entries such that 
 \[
  \Nout R  - H(LY^{\Nout}) + I(LY^{\Nout};D^{\Nout},B)
 \]
  vanishes
 exponentially in $\Nout$.
 \label{lemma:secretkey}
\end{lemma}
In other words, $LY^{\Nout}$ is an $\epsilon$-SK for suitable $\epsilon$.
 Let $q=p^k$ and $B=(\wv,s_{\tv}^{\Nout}:\tv\in V^*)$. Then, the above lemma guarantees the existence of an $\mathbb{F}_{p^k}$-valued matrix $L$, so that $L(y^{(1)},\ldots , y^{(\Nout)})^T$ is a secret key with a rate of
 \begin{equation}
  \Rkey = \frac{1}{N}H(\yv:\tv\in V^*|\dv:\tv\in V^*) - \sum_{\tv\in V^*}\Rcomv, 
  \label{eq:rkey_ach}
 \end{equation}
 where $\Rcomv$ denotes the total rate of communication of terminal $\tv$.
We give a lower bound on $\Rkey$ by bounding $H(\yv:\tv\in V^*|\dv:\tv\in V^*)$ in the next section.

\subsection{Joint Entropy of the Quantized Sources}
The proof of the following lemma is given in Appendix~\ref{sec:proof_lemmajointentropy}.
\begin{lemma}
Fix a $\delta>0$, and let $D_i:=(\dvi:\tv\in V^*)$. For all sufficiently large $n$, we have
\begin{align}
&\frac{1}{n}H(\yvi:\tv\in V^*|D_i) \geq  \frac{1}{2}\log_2\left( 1+\frac{1}{\sigma^2(\L_{\rTst})}\right) &\notag\\
                   &\hspace{2cm} +\sum_{\tu\in V^*} \frac{1}{2}\log_2\left(1+ \frac{{1-\rho^2_{\tu, \parent(\tu)}}}{\sigma^2(\Lu)} \right)-\delta \label{eq:joint_entropy}
\end{align}
\label{lemma:jointentropy}
\end{lemma}
For every $\tv$, $\{\yvi:1\leq i\leq \Nout\}$ are independent and identically distributed. If $D:=\{ D_i: 1\leq i\leq \Nout \}$, then $H(\yv:\tv\in V^*|D)=\Nout H(\yvi:\tv\in V^*|D_i)$. Substituting for $\sigma^2(\Lv)$ from (\ref{eq:quantizationrate}) in (\ref{eq:joint_entropy}), we get $$\frac{1}{N}H(\yv:\tv\in V^*|D)\geq \mathcal{R}_{\mathrm{ent}}-g(\delta)-o_n(1),$$ where $\mathcal{R}_{\mathrm{ent}}$ is defined in (\ref{eq:Rent}), and $g(\delta)$ is a quantity that goes to $0$ as $\delta\to 0$.

\subsection{Achievable Secret Key Rate and Proof of Theorem~\ref{thm:main}}
Lemma~\ref{lemma:secretkey} guarantees the existence of a strong secret key which is a linear transformation of $(y^{(1)},\ldots,y^{(\Nout)})$.
From Propositions~\ref{prop:Pe_analog} and~\ref{prop:Pe_digital}, all terminals are able to recover $(y^{(1)},\ldots,y^{(\Nout)})$ with a 
probability of error that decays exponentially in $N=n\Nout$. 

During the analog phase, each terminal $\tv$ in $V^*$ publicly communicates 
\begin{align}
	R_{\mathrm{analog}}^{(\tv)}&=\frac{1}{n}\log_2\frac{\vol(\Lva)}{\vol(\Lv)}&\notag\\
	      &\leq \max_{\tu\in N_{T}(\tv)}\frac{1}{2}\log_2 \frac{\left( {1-\rhouv^2} +\sigma^2(\Lv)+\rhouv^2\sigma^2(\Lu)  \right)}{\sigma^2(\Lv)}&\notag\\
	       &\qquad \qquad\qquad +o_n(1)+\delta.
\end{align}
bits per sample. Here, we have used the fact that an MSE quantization-good lattice $\Lv$ satisfies $\vol(\Lv)\to 2\pi e\sigma^2(\L)$ as $n\to\infty$. We know from (\ref{eq:digital_leakage}) that during the digital phase, terminal $\tv$ communicates $2\delta \Rqv$ bits per sample across the public channel.
The total rate of communication by terminal $\tv$ is therefore
\begin{align}
\Rcomv &\leq \max_{\tu\in N_{T}(\tv)}\frac{1}{2}\log_2 \frac{\left( {1-\rhouv^2} +\sigma^2(\Lv)+\rhouv^2\sigma^2(\Lu)  \right)}{\sigma^2(\Lv)}&\notag\\
       &\hspace{3cm} +\delta(1+2\Rqv)+o_n(1) &\label{eq:total_commrate}
\end{align}
bits per sample. Using Lemma~\ref{lemma:jointentropy} and (\ref{eq:total_commrate}) in (\ref{eq:rkey_ach}), and finally substituting  (\ref{eq:quantizationrate}), we obtain (\ref{eq:ach_SKrates}).
All that remains now is to find an upper bound on the computational complexity of our scheme.

\subsection{Computation Complexity}
We now show that the computational complexity is polynomial in the number of samples $N$.
The complexity is measured in terms of the number of binary operations required, and we make the assumption that  
each floating-point operation (i.e., operations in $\R$) requires $O(1)$ binary operations.
In other words, the complexity of a floating-point operation is independent of $N$.

Recall that $N=n\Nout$, where $\Nout=\min_{\tv\in V^*}(2^{n\Rqv}-1)$. Also, $\alpha =(\max_{\tv\in V^*}\Rqv)/(\min_{\tv\in V^*}\Rqv)$.

\begin{itemize}
\item \emph{Quantization}: Each lattice quantization operation has complexity at most $O(2^{n\Rqv})=O(\Nout^{\alpha})$. There are $\Nout$ such quantization operations to be performed at each terminal, and hence the total complexity is at most $O(\Nout^{\alpha+1})$.
\item \emph{Analog Phase}: Terminal $\tv$ performs $\Nout$ quantization and $\bmod\Lva$ operations to compute $\{\wvi:1\leq i\leq\Nout \}$, and this requires a total complexity of $O(\Nout^{\alpha+1})$. Computation of $\{\hyvi: 1\leq i\leq\Nout, \tv\in V^* \}$ requires at most $\Nout(|V^*|-1)$ quantization operations, which also has a total complexity of $O(\Nout^{\alpha+1})$.
\item \emph{Digital Phase}: Each terminal has to compute the coset representative. This is followed by the decoding of the Reed-Solomon code. Both can be done using the Reed-Solomon decoder, and this requires $O(\Nout\log_2\Nout)$ operations in $\Fpkv$. Each finite field operation on the other hand requires $O(\log_2^2p^{k_{\tv}})=O(n^2)$ binary operations~\cite[Chapter 2]{Hankersonbook}. The total complexity is therefore $O(N^2)$.
\item \emph{Secret Key Generation}: This involves multiplication of a $\lfloor \frac{\Nout R}{\log_2 q} \rfloor\times \Nout$ matrix with an $\Nout$-length vector, which requires $O(\Nout^2/\log q)$ operations over $\mathbb{F}_q$. Hence, the complexity required is $O(\Nout^2\log q)=O(N^2)$.
\end{itemize}
From all of the above, we can conclude that the complexity required is at most $O(N^{\alpha +1})$. If the quantization rate constraints are the same, i.e., $\Rqu=\Rqv$, then the complexity is $O(N^2)$. This completes the proof of Theorem~\ref{thm:main}.
\qed

\section{Acknowledgments}
The first author would like to thank Manuj Mukherjee for useful discussions. The work of the first author was supported by the TCS Research scholarship programme, and that of the second author by a Swarnajayanti fellowship awarded by the Department of Science and Technology (DST), India.

\begin{appendices}
\section{Lattice Concepts}\label{sec:latticeconcepts}

In this appendix, we briefly review basic lattice concepts that are relevant to this work. We direct the interested reader to~\cite{Conway,Erez04,Erez05,Zamirbook} for more details.
Let $A$ denote a full-rank $n\times n$ matrix with real entries. Then the set of all integer-linear combinations of the columns of $A$ is called a lattice in $\R^n$, and $A$ is called a generator matrix of the lattice. 
Given a lattice $\L$ in $\R^n$, we define $Q_{\L}:\R^n\to \L$ to be the lattice quantizer that maps every point in $\R^n$ to the closest (in terms of Euclidean distance) point in $\L$, with ties being resolved according to a fixed rule. The \emph{fundamental Voronoi region}, $\cV(\L)$, is the set of all points in $\R^n$ for which $\0$ is the closest lattice point, i.e., $\cV(\L):=\{\x\in \R^n:Q_{\L}(\x)=\0\}$. For any $\x\in\R^n$, we define $[\x]\bmod\L:=\x-Q_{\L}(\x)$. We also define $\vol (\L):= \vol(\cV(\L))$. The \emph{covering radius} of $\L$, denoted $\rcov(\L)$ is the radius of the smallest closed ball in $\R^n$ centered at $\0$ that contains $\cV(\L)$. Similarly, the \emph{effective radius}, $\reff(\L)$, is the radius of a ball in $\R^n$ having volume $\vol(\L)$.

The \emph{second moment per dimension} of a lattice, $\sigma^2(\L)$ is defined as 
\[
	\sigma^2(\L) = \frac{1}{n\vol(\L)}\int_{\x\in\cV(\L)}\Vert \x\Vert^2 d\x,
\]
 and is equal to the second moment per dimension of a random vector uniformly distributed over $\cV(\L)$. The \emph{normalized second moment per dimension} of $\L$
 is defined as 
 \[
 	G(\L):=\frac{\sigma^2(\L)}{\vol(\L)^{2/n}}.
 \]
 
 If $\L$ and $\Lc$ are two lattices in $\R^n$ that satisfy $\Lc\subset \L$, then we say that $\Lc$ is a \emph{sublattice} of $\L$, or $\Lc$ is \emph{nested} in $\L$. Furthermore,
 \[
 	|\L\cap\cV(\Lc)| = \frac{\vol(\Lc)}{\vol(\L)}.
 \]
 
 We say that a lattice $\L$ (or more precisely, a sequence of lattices $\{\L \}$ indexed by the dimension $n$) is \emph{good for mean squared error (MSE) quantization} if
 \[
 	\lim_{n\to\infty}G(\L) =\frac{1}{2\pi e}. 
 \]
 A useful property is that if $\L$ is good for MSE quantization, then $\vol(\L)^{2/n}\to 2\pi e \sigma^2(\L)$ as $n\to\infty$.
 We say that $\L$ is \emph{good for covering} (or covering-good or Rogers-good) if $\rcov(\L)/\reff(\L)\to 1$ as $n\to\infty$. It is a fact that if $\L$ is good for covering, then it is also good for MSE quantization~\cite{Erez05}.
%
%

Let $\bZ$ be a zero-mean $n$-dimensional white Gaussian vector having second moment per dimension equal to $\nsvar$. Let
\[
{      \mu:=\frac{\vol\big( \cV(\L) \big)^{2/n}}{\sigma^{2}}  }.
\]
 Then we say that $\{\L\}$ is \emph{good for AWGN channel coding} (or AWGN-good or Poltyrev-good) if the probability that $\bZ$ lies outside the fundamental Voronoi region of $\L$ is upper bounded by
\[
\Pr[ \bZ \notin \cV(\L) ] \leq e^{-n\big( E_{U}(\mu)-o_{n}(1) \big)}
\]
for all $\nsvar$ that satisfy $\mu\geq 2\pi e$.
Here, $E_{U}(\cdot)$, called the \emph{Poltyrev exponent} is defined as follows:
\begin{equation}
E_{U}(\mu) = \begin{cases}
\frac{\mu}{16\pi e} & \text{ if } 8\pi e \leq \mu \\
\frac{1}{2}\ln\frac{\mu}{8\pi} &\text{ if } 4\pi e \leq \mu \leq 8\pi e \\
\frac{\mu}{4\pi e} - \frac{1}{2}\ln\frac{\mu}{2\pi} &\text{ if } 2\pi e \leq \mu \leq 4\pi e.
\end{cases}
\label{eq:polty_exp}
\end{equation}
Suppose that we use a subcollection of points from an AWGN-good lattice $\L$ as the codebook for transmission over an AWGN channel. Then, as long as 
\[
\frac{\vol\big( \cV(\L) \big)^{2/n}}{\sigma^{2}} \geq 2\pi e,
\] 
the probability that a lattice decoder decodes to a lattice point other than the one that was transmitted, decays exponentially in the dimension $d$, with the exponent given by (\ref{eq:polty_exp}). 

Lattices that satisfy the above ``goodness'' properties were shown to exist in~\cite{Erez05}. Moreover, such lattice can be constructed from linear codes over prime fields. Let $p$ be a prime number, and $\cC$ be an $(n,k)$ linear code over $\mathbb{F}_p$. In other words, $\cC$ has blocklength $n$ and dimension $k$.
Let $\phi$ be the natural embedding of $\mathbb{F}_p$ in $\Z$, and for any $\x\in \mathbb{F}_p^n$, let $\phi(\x)$ be the $n$-length vector obtained by operating $\phi$ on each component of $\x$. The set $\L := \phi(\cC)+p\Z^n:=\{ \phi(\x)+p\y: \x\in\cC,\y\in \Z^n  \}$ is a lattice, and is called the Construction-A lattice obtained from the linear code $\cC$. With a slight abuse of notation, we will call any scaled version of $\L$, i.e., $\alpha \L$ for any $\alpha>0$, a \emph{Construction-A} lattice obtained from $\cC$. A useful fact is that $\L$ always contains $p\Z^n$ as a sublattice, and the nesting ratio $\L/pZ^n=p^{k}$. It was shown in~\cite{Erez05} that if $k$ and $p$ are appropriately chosen functions of $n$, then a randomly chosen Construction-A lattice over $\mathbb{F}_p$ is good for covering and AWGN channel coding with probability tending to $1$ as $n\to\infty$.

We use the nested lattice construction in~\cite{Erez04,Krithivasan_goodlattice} to obtain good nested lattices. 
Let $\Lc$ be a Construction-A lattice which is good for covering and AWGN, and let $A$ be a generator matrix for $\Lc$. Then, if $\L'$ is another Construction-A lattice, then $\L=p^{-1}A\L'$ is a lattice that contains $\Lc$ as a sublattice. It was shown in~\cite{Krithivasan_goodlattice} that if $\L$ and $\Lc$ are chosen at random, then they are both simultaneously good for AWGN channel coding and covering with probability tending to $1$ as $n\to\infty$ (provided that $k$ and $p$ are suitably chosen).


\section{Technical Proofs}

\subsection{Proof of Proposition~\ref{prop:Pe_analog}}\label{sec:proof_prop_analog}
Recall that 
\begin{align}
	\yui &= [Q_{\Lu}(\xui+\dui)-\dui]\bmod\Lub&\notag\\
	     &= \left[\xui-[\xui+\dui]\bmod\Lu\right] \bmod\Lub &\notag\\
	     &= [\xui+\tdui]\bmod\Lub,
\end{align}
where $\tdui$ is uniformly distributed over $\cV(\Lu)$ and is independent of $\xui$~\cite[Lemma 1]{Erez04}.
Since $\Lu$ is good for MSE quantization, $\Lub$ is good for AWGN and (\ref{eq:volLub_constraint}) is satisfied, we can use~\cite[Theorem 4]{Erez05} to assert that\footnote{Note that there is a slight difference here since $\tdui$ is not Gaussian. However, the arguments in~\cite[Theorem 5]{Erez04} can be used to show that $\xui+\tdui$ can be approximated by a Gaussian since $\Lu$ is good for MSE quantization.} the probability
\begin{equation}
	\Pr[\yui\neq \xui+\tdui] \leq e^{-n(E_1(\delta)-o_n(1))}
	\label{eq:appendix1_1}
\end{equation}
where $E_1(\delta)>0$ for all $\delta>0$. Similarly, we can write
\[
 \yvi = [\xvi+\tdvi]\bmod\Lvb,
\]
where $\tdvi$ is independent of $\xvi$, and 
\begin{equation}
	\Pr[\yvi\neq \xvi+\tdvi] \leq e^{-n(E_2(\delta)-o_n(1))}
	\label{eq:appendix1_2}
\end{equation}
where $E_2(\delta)>0$ for all $\delta>0$.

Recall that $\wvi=[\yvi]\bmod\Lva$.
We can write
\begin{align}
\hyvi &= \wvi + Q_{\Lva}(\rhouv \yui-\wvi)&\notag \\
      &= \wvi + Q_{\Lva}\left(\rhouv\yui -\yvi+Q_{\Lva}(\yvi)\right) &\notag\\
      &= \wvi +Q_{\Lva}(\yvi)+ Q_{\Lva}\left(\rhouv\yui -\yvi\right)&\notag \\
      &= \yvi + Q_{\Lva}\left(\rhouv\yui -\yvi\right)&\label{eq:appendix1_3}
\end{align}
From (\ref{eq:appendix1_1}) and (\ref{eq:appendix1_2}), we know that $\yvi=\xvi+\tdvi$ and $\yui=\xui+\tdui$ with high probability. 
Now, 
\[
	\rhouv\xui -\xvi =-\sqrt{1-\rhouv^2}\:\zvui, 
\]
and again using the AWGN goodness property of $\Lva$ and (\ref{eq:volLva_constraint}), we have
\begin{multline}
 \Pr\left[Q_{\Lva}(-\sqrt{1-\rhouv^2}\:\zvui+\tdui-\tdvi)\neq \0 \right] \\ \leq e^{-n(E_{3}(\delta)-o_n(1))}
\label{eq:appendix1_4}
\end{multline}
where $E_3(\delta)>0$ for $\delta>0$. Using (\ref{eq:appendix1_1}), (\ref{eq:appendix1_2}) and (\ref{eq:appendix1_4}), we get that
\[
	\Pr[\hyvi\neq  \yvi] \leq \sum_{i=1}^{3}e^{-n(E_i(\delta)-o_n(1))}
\]
which completes the proof of the proposition.
\qed

\subsection{Proof of Lemma~\ref{lemma:jointentropy}}\label{sec:proof_lemmajointentropy}
We prove the result by expanding the joint entropy using the chain rule, and then use a lower bound on the entropy of a quantized Gaussian.
To do this, we will expand the joint entropy in a particular order. Let $\mathcal{S}$ be any (totally) ordered set containing the vertices of $T^*$ and 
satisfying the following properties:
\begin{itemize}
\item $\max_{\tv}\mathcal{S}=\rTst $, i.e., $\rTst\geq \tv$ for all $\tv\in \mathcal{S}$.
\item $\tv>\tu$ if the distance between $\tv$ and $\rTst$ is less than that between $\tu$ and $\rTst$.
\end{itemize}
Essentially, $\tv>\tu$ if $\tv$ is closer to $\rTst$ than $\tu$, and we do not care how the vertices at the same level (vertices at the same distance from $\rTst$) are ordered. 
Let $D=(\dvi:\tv\in V^*)$. Then,
\begin{align}
H(\yvi:\tv\in V^*|D) &= H(\y_{\rTst}^{(i)}|D)&\notag \\
				   &\quad + \sum_{\tv\in V^*\backslash \rTst} H(\yvi|D,\yui:\tu>\tv)&\notag\\
                   &\geq H(\y_{\rTst}^{(i)}|D)&\notag \\
                   &\quad + \sum_{\tv\in V^*\backslash \rTst} H(\yvi|D,\xui:\tu>\tv)&\label{eq:prf_lem_entropy05}\\
                   &= H(\y_{\rTst}^{(i)}|D)&\notag \\
                   &\quad + \sum_{\tv\in V^*\backslash \rTst} H(\yvi|D,\x_{\parent(\tv)}^{(i)})&\label{eq:prf_lem_entropy_1}
\end{align}
where  (\ref{eq:prf_lem_entropy05}) follows from the data processing inequality. 
We would like to remark that (\ref{eq:prf_lem_entropy_1}) is the only place where we use Markov tree assumption.
The rest of the proof closely follows \cite[Lemma 4.3]{Nitinawarat12}, and we give an outline.
The idea is to find the average mean squared error distortion in representing $\xvi$ by $\yvi$ (with or without the side information $\x_{\parent(\tv)}^{(i)}$),
and then argue that the rate of such a quantizer must be greater than or equal to the rate-distortion function.

\begin{claim}
\begin{equation}
 \frac{1}{n}H(\y_{\rTst}^{(i)}|D) \geq \frac{1}{2}\log_{2}\left( 1+\frac{1}{\sigma^2(\L_{\rTst})} \right)-o_n(1).
 \label{eq:entropy_firstterm}
\end{equation}
\label{claim:1}
\end{claim}
Making minor modifications to the proof of~\cite[Lemma 4.3]{Nitinawarat12}, we can show that conditioned on $D$, the average MSE distortion (averaged over $D$) between $\x_{\rTst}^{(i)}$ and $$\widehat{\x}_{\rTst}^{(i)}=\frac{1}{1+\sigma^2(\L_{\rTst})}\y_{\rTst}^{(i)}$$ is at most $\frac{\sigma^2(\L_{\rTst})}{1+\sigma^2(\L_{\rTst})}+o_n(1)$.
Since any rate-distortion code for quantizing $\x_{\rTst}^{(i)}$ must have a rate at least as much as the rate-distortion function, we can show that (again following the proof of~\cite[Lemma 4.3]{Nitinawarat12}) Claim~\ref{claim:1} is true.

\begin{claim}
\begin{equation}
\frac{1}{n}H(\yvi|D,\x_{\parent(\tv)}^{(i)}) \geq \frac{1}{2}\log_2\left( 1+\frac{{1-\rho^2_{\tv,\parent(\tv)}}}{\sigma^2(\Lv)} \right)-o_n(1)
\label{eq:entropy_otherterms}
\end{equation}
\end{claim}
 The proof of the above claim also follows the same technique. We can show that conditioned on $D$ and $\x_{\parent(\tv)}^{(i)}$, the average MSE distortion between $\sqrt{1-\rhouv^2}\zvui$ and 
 \[
 \widehat{\z}_{\tv\tu}^{(i)} = \frac{(1-\rhouv^2)}{1-\rhouv^2+\sigma^2(\Lv)}\left[\yvi-\rhouv\x_{\parent(\tv)}^{(i)}\right]\bmod\Lvb
 \]
 is $\frac{(1-\rhouv^2)\sigma^2(\Lv)}{1-\rhouv^2+\sigma^2(\Lv)}+o_n(1)$. Arguing as before, the claim follows.
 
 Finally, using (\ref{eq:entropy_firstterm}) and (\ref{eq:entropy_otherterms}) in (\ref{eq:prf_lem_entropy_1}) completes the proof of Lemma~\ref{lemma:jointentropy}.
 \qed

\end{appendices}

\end{document}